\documentstyle[pramana,epsf,epsfig,floats]{ias}
\begin{document}
\mark{{Searching for Quark Matter...}{Itzhak Tserruya}}
\title{Searching for Quark Matter with Dileptons and Photons:
               from SPS to RHIC}

\author{Itzhak Tserruya \footnote{e-mail: Itzhak.Tserruya@weizmann.ac.il.\\ 
                         Work supported by the Israeli Science Foundation, the MINERVA
                         Foundation and the US-Israel Binational Science Foundation.}}

\address{Weizmann Institute of Science, \\
         Rehovot 76100, Israel}


\abstract{
  
  The heavy-ion programme at the CERN SPS, which started back in '86, has produced a wealth 
of very interesting and intriguing results in the quest for the quark-gluon plasma.
The highlights of the programme on dilepton and direct photon measurements 
are reviewed emphasizing the most recent results obtained in Pb-Pb 
collisions at 158 A GeV. Prospects from RHIC are discussed.
}
\maketitle

\section{INTRODUCTION}
 The heavy-ion programme at the CERN SPS started in 1986 with the acceleration of 
O beams at an energy of 200 GeV/c per nucleon, later followed 
by a S beam at the same energy. Since 1994 the experiments used a Pb beam at a maximum energy 
of 158 GeV/c per nucleon ($\sqrt{s_{NN}}$=17.2 GeV/c). The primary scientific objective is the search for
the predicted phase transition associated with quark-gluon plasma formation and chiral symmetry 
restoration, using relativistic nuclear collisions to achieve in the laboratory the necessary 
conditions of temperature and density.

  The CERN programme has produced a wealth of very interesting and intriguing results in the quest 
for the quark-gluon plasma.  
From systematic studies of global observables, particle distributions, production rates 
and correlations, a picture  of a chemically and thermally equilibrated system 
undergoing  collective expansion has emerged. 
The observation of J/$\psi$ suppression, strangeness enhancement and excess emission of 
low-mass lepton pairs are among the most notable results hinting at new physics \cite{qm}. 
   
 The Relativistic Heavy-Ion Collider (RHIC) at BNL started regular operations in the summer
of 2000, opening a new era in the study of ultrarelativistic heavy-ion collisions, 
at energies up to $\sqrt{s_{NN}}$=200 GeV/c, more than one order of magnitude larger than at CERN. 
The higher c.m. energies result in much better conditions 
(higher energy density, higher temperature, longer lifetime) for the study of quark matter.

The advent of RHIC seems to be an appropriate time to summarize the achievements at CERN  
and to put them in perspective with respect to the expectations from RHIC.
The discussion in this paper is restricted to electromagnetic probes, dilepton and photon
measurements, only.
 
   The interest in dileptons and photons stems from 
their relatively large mean free path. As a consequence, they
can leave the interaction region without final state interaction, carrying information about 
the conditions and properties of the matter at the time of their production
and in particular of the early stages of the collision when temperature and energy density have 
their largest values, i.e. when the conjectured deconfinement and chiral symmetry restoration phase 
transition is expected to take place. 
This has to be contrasted with the hadronic observables which are sensitive to the late stages
at, or after freeze-out, when the hadronic system stops interacting.
 
  A prominent topic of interest is the identification of thermal radiation emitted from 
the collision system. This radiation should tell us the nature of the matter formed, 
a quark-gluon plasma (QGP) or a high-density hadron gas (HG).  

The physics potential of dileptons is further emphasized by the capability to measure 
the vector mesons $\rho,\omega$ and $\phi$, through their leptonic decays. 
Of particular interest is the $\rho$ meson since it provides a unique 
opportunity to observe in-medium modifications of the vector meson properties (mass and/or width)
which might be linked to chiral symmetry restoration. 
   Due to its very short lifetime  ($\tau$ = 1.3 fm/c) compared to the typical fireball 
   lifetime of $\sim$10 fm/c at SPS energies, most of the $\rho$ mesons 
   decay inside the interaction region with eventually modified properties.
   The situation is  different for the $\omega$ and $\phi$ mesons. Because 
   of their much longer lifetimes they predominantly decay 
   outside the interaction region after having regained their vacuum properties.

The CERN experiments have confirmed the unique physics potential of these electromagnetic probes. A 
compilation of all measurements performed so far is presented in Table ~1 together with the kinematic 
phase space covered and relevant references. The programme involved systematic studies with p, S and 
Pb beams and the most notable results will be reviewed below. 
\begin{table}[h!]
%\vspace{-0.6cm} 
\begin{center}
{ Table 1. List of Measurements at the CERN SPS \\
                DILEPTONS }\\[0.3cm]
\leavevmode
\vbox{\columnwidth=12cm
\begin{tabular}{|c|c|c|c|c|c|}
%\hline\hline
 Experiment &  Probe      &  System           &  $y$      &    Mass       & Ref. \\
            &             &                   &           & (GeV/c$^2$)   &      \\ \hline
            &             & p-Be,Au 450 GeV/c &           &               & 2,3  \\ 
 CERES      &   $e^+e^-$  & S-Au    200 GeV/u &  2.1-2.65 & 0 -- 1.4      & 4    \\
            &             & Pb-Au   158 GeV/u &           &               & 5-7  \\
            &             &          40 GeV/u &           &               & 8    \\[0.2cm]
  \hline 
 HELIOS-1   &$\mu^+\mu^-$ & p-Be 450 GeV/c    & 3.65-4.9  & 0.3 -- 4.0    & 9    \\
 (completed)& $e^+e^-$    &     ``            & 3.15-4.65 &      ``       & 9   \\[0.2cm]
\hline
 HELIOS-3   &$\mu^+\mu^-$ & p-W,S-W 200 GeV/u & $>3.5$    & 0.3 -- 4.0    & 10   \\
 (completed)&             &                   &           &               &      \\[0.2cm]
\hline
 NA38       & $\mu^+\mu^-$& p-A,S-U 200 GeV/u & 3.0-4.0   & 0.3 -- 6.     & 11   \\
 NA50       &     ``      & Pb-Pb   158 GeV/u &    ``     & 0.3 -- 7.0    & 12   \\[0.2cm]
%\hline\hline
\end{tabular}
}
\end{center}
\end{table}
\begin{table}[h!]
%\vspace{-2.0cm}
\begin{center}
{PHOTONS} \\[0.3cm]
\leavevmode
\vbox{\columnwidth=10cm
\begin{tabular}{|c|c|c|c|c|}
%\hline\hline
 Experiment    &  System           &   $y$      &  $p_t$ (GeV/c)  & Ref.  \\ [0.2cm] \hline 
CERES          &  S-Au 200 GeV/u   &   2.1-2.7  &    0.4-2.0      & 13  \\
               &  Pb-Au 158 GeV/u  &     ``     &      ``         &  \\ [0.2cm] \hline 
HELIOS-2       & p,O,S-W 200 GeV/u &   1.0-1.9  &    0.1-1.5      & 14  \\ 
(completed)    &                   &            &                 &  \\[0.2cm] \hline 
WA80/ WA98     &  O-Au   200GeV/u  &   1.5-2.1  &    0.4-2.8      &15  \\
(completed)    &  S-Au   200 GeV/u &   2.1-2.9  &    0.5-2.5      & 16 \\
               & Pb-Au   158 GeV/c &       ``   &       ``        & 17  \\ [0.2cm]
%\hline \hline
\end{tabular}
}
\end{center}
\end{table}   
%\vspace{-1.2cm}

\section{LOW-MASS DILEPTONS}
\subsection{Experimental Results}
  The low-mass region, m = 200 - 600 MeV/c$^2$, has been systematically studied by the CERES 
experiment including measurements of p-Be (a very good approximation to pp collisions) and
p-Au at 450 GeV/c 
\cite{pbe-ee,pbe-eegamma}, S-Au at 200 GeV per nucleon \cite{prl95},
and Pb-Au at 158 GeV per nucleon \cite{ir-qm97,pbau95-plb,bl-qm99} and 
40 GeV per nucleon \cite{pb-au40}. The Pb-Au results at 158 GeV per nucleon
were obtained in two different runs, in 1995
\cite{ir-qm97,pbau95-plb} and 1996 \cite{bl-qm99}.
Apart from a slight difference in the centrality trigger ($<$dn$_{ch}$/d$\eta>$ = 250 and 220
in the '96 and '95 runs, respectively), the two measurements were performed under identical
conditions and yielded consistent results within their systematic uncertainties. 
Fig. 1 shows the dilepton invariant mass spectrum of the '96 Pb-Au run at 158 GeV per nucleon
as well as the preliminary results obtained in the '99 Pb-Au run at 40 GeV per nucleon. In both cases 
the dilepton yield is normalized to the measured charged particle yield within the CERES acceptance 
($2.1 < \eta <2.65$). 
%%%%FIGURE 1. CERES RESULTS-- 96 Pb DATA + PRELIMINARY 40 GEV DATA  WITH HI-THERMAL COCKTAIL  
\begin{figure}[h!]
%\vspace{-1.7cm}
\begin{minipage}[t]{65mm}
\epsfig{file=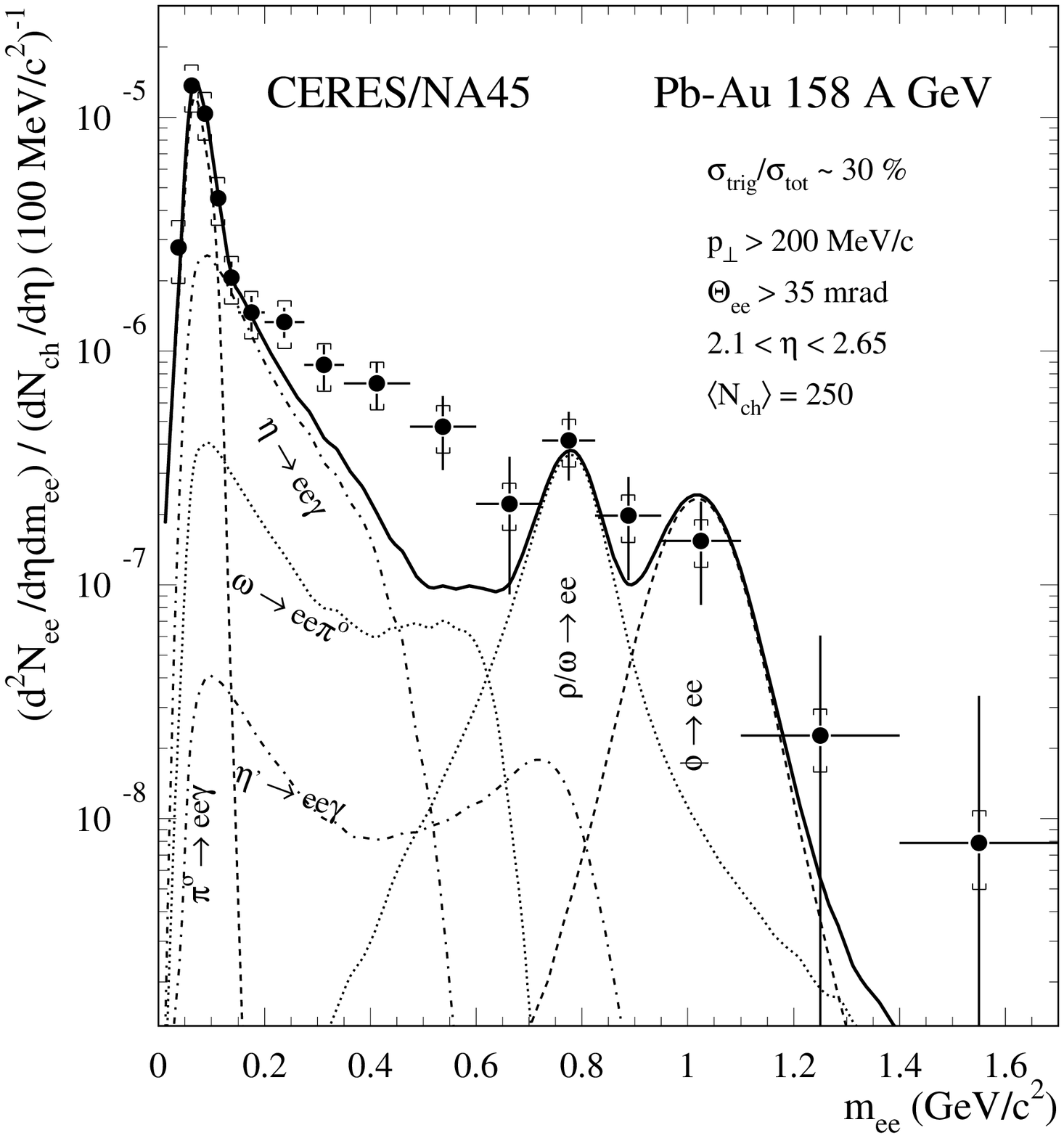,width=6.5cm,height=7.0cm}
\end{minipage}
\begin{minipage}[h!]{65mm}
\vspace*{-6.8cm}
\hspace*{6.2cm}
\epsfig{file=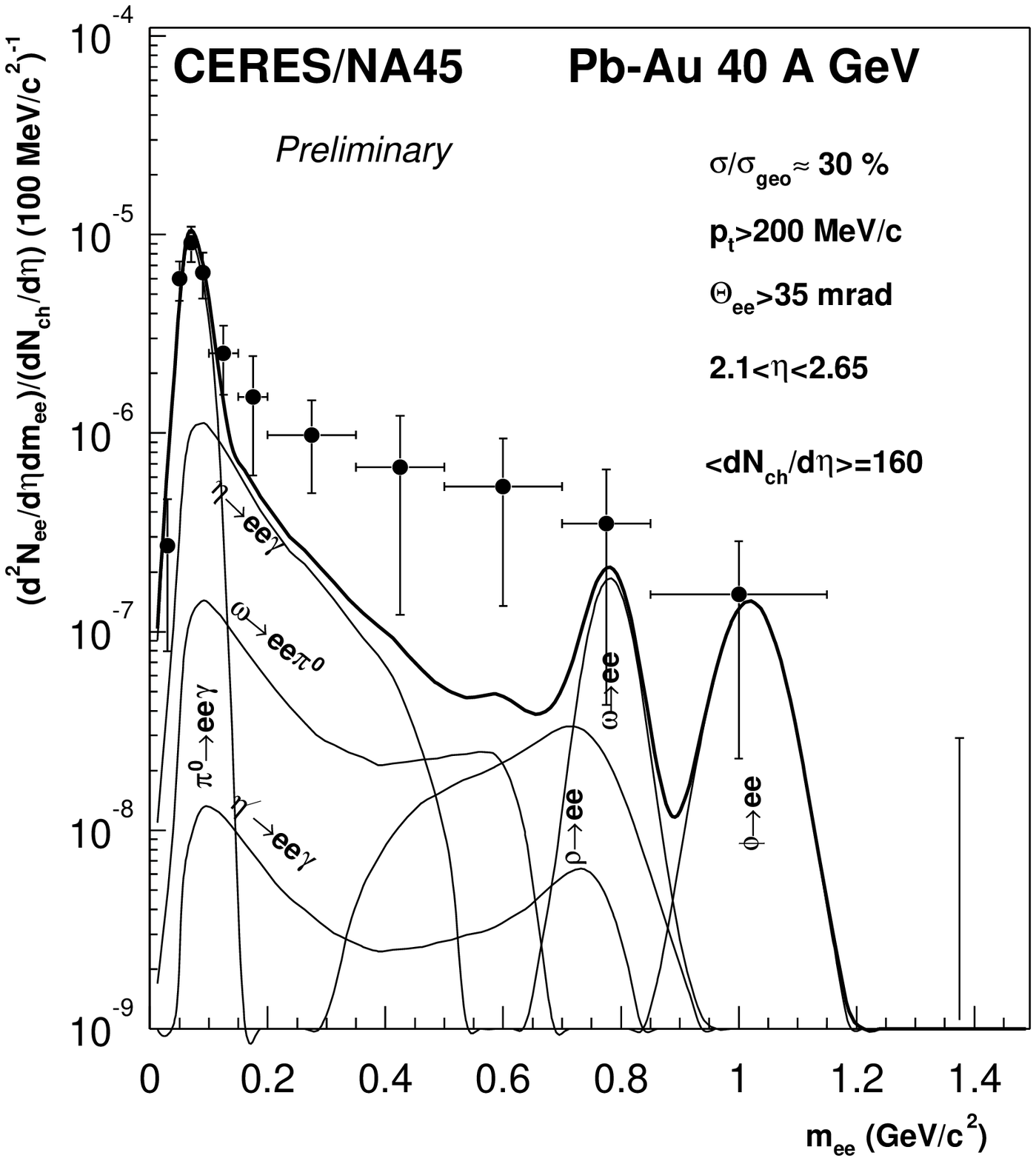,width=6.5cm,height=6.5cm}
\end{minipage}
\caption {Inclusive $e^+e^-$ mass spectrum measured by CERES in 158 A~GeV Pb--Au 
         collisions in the '96 run (left panel)~\protect\cite{bl-qm99} and 40 A~GeV (right panel)
         ~\protect\cite{pb-au40}. The figures also show the 
         summed (thick solid line) and individual (dotted lines in the left panel and thin lines in the
          right panel) contributions from hadronic sources.}
%\vspace{-0.7cm}
\end{figure}
The solid lines represent the expected yield from known hadronic sources based on a generator 
\cite{bl-qm99} which uses measured particle production ratios whenever available or ratios calculated
with a thermal model which describes well all measured ratios \cite{pbm-js} 
\footnote{ The standard pp cocktail, previously used in the 
presentation of the CERES results \cite{pbe-ee} and based on yields directly measured in pp
collisions, predicts very similar results. The total yield of the present generator is
$\sim$30 \% larger than the pp cocktail for masses m $>$ 200 MeV/c$^2$, the main difference occurring 
in the region of the $\phi$ meson.}. 
As previously observed with the S beam \cite{prl95}, the low-mass continuum in Pb-Au collisions
is strongly enhanced with respect to this cocktail, both at 158 GeV per nucleon and at the much 
lower energy of 40 GeV per nucleon.
For example, in the mass region  m = 0.25 - 0.7 GeV/c$^2$, the enhancement factor (defined as the
ratio of the measured to the calculated yield) at 158 GeV 
per nucleon is 2.6 $\pm$0.5(stat.) $\pm$0.6(syst.).

 CERES has further characterized the properties of the low-mass excess. The results indicate that 
it is mainly due to pairs with soft p$_t$ and that it increases
faster than linearly with the event multiplicity \cite{ir-qm97,pbau95-plb,bl-qm99}. The latter point is
illustrated in Fig.~2 which displays the enhancement factor as function of the charged particle
rapidity density for four mass intervals. Since the calculated yield increases linearly with multiplicity, 
a constant enhancement factor indicates that the data also increases linearly with multiplicity, as is the
case for the very low-masses, in the region of the $\pi^0$ Dalitz decay. However, in the mass
interval $0.25 < m < 0.68$~GeV/c$^2$ the measured yield increases stronger than linearly.
%%% FIGURE 2 CERES 96 RESULTS. MULTIPLICITY DEPENDENCE 
\begin{figure}[h!]
%\vspace{-2.0cm}
\centerline{\epsfxsize=8.0cm \epsfbox{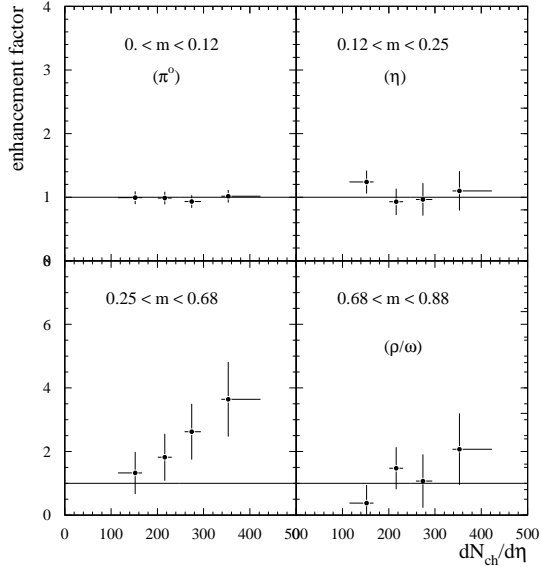}}
%\vspace{-1.8cm}
\caption {Multiplicity dependence of the enhancement factor for four mass intervals  
          obtained by CERES in Pb-Au collisions at 158 GeV per nucleon in the '96 run ~\protect\cite{bl-qm99}.}
%\vspace{-0.5cm}
\end{figure}

   An enhancement of low-mass dileptons has also been observed in the di-muon experiments
\cite{helios-3,falco-qm97,bordalo-qm99}. Whereas the p-U data of NA38  
are well reproduced by a cocktail of hadronic sources (with the somewhat uncertain extrapolation of the 
Drell-Yan contribution into low masses), the S  
data show an enhancement of low-mass pairs which is hardly noticeable in the Pb-Pb collisions
(see Fig. 3). In both cases the enhancement is clearly visible in the $\phi$ meson and extends 
%%%%%% FIGURE 3. NA38 S-U  MASS SPECTRUM AND NA50 PB-PB SPECTRUM
\begin{figure}[t]
%\vspace{-1.0cm}
\begin{minipage}[t]{65mm}
\epsfig{file=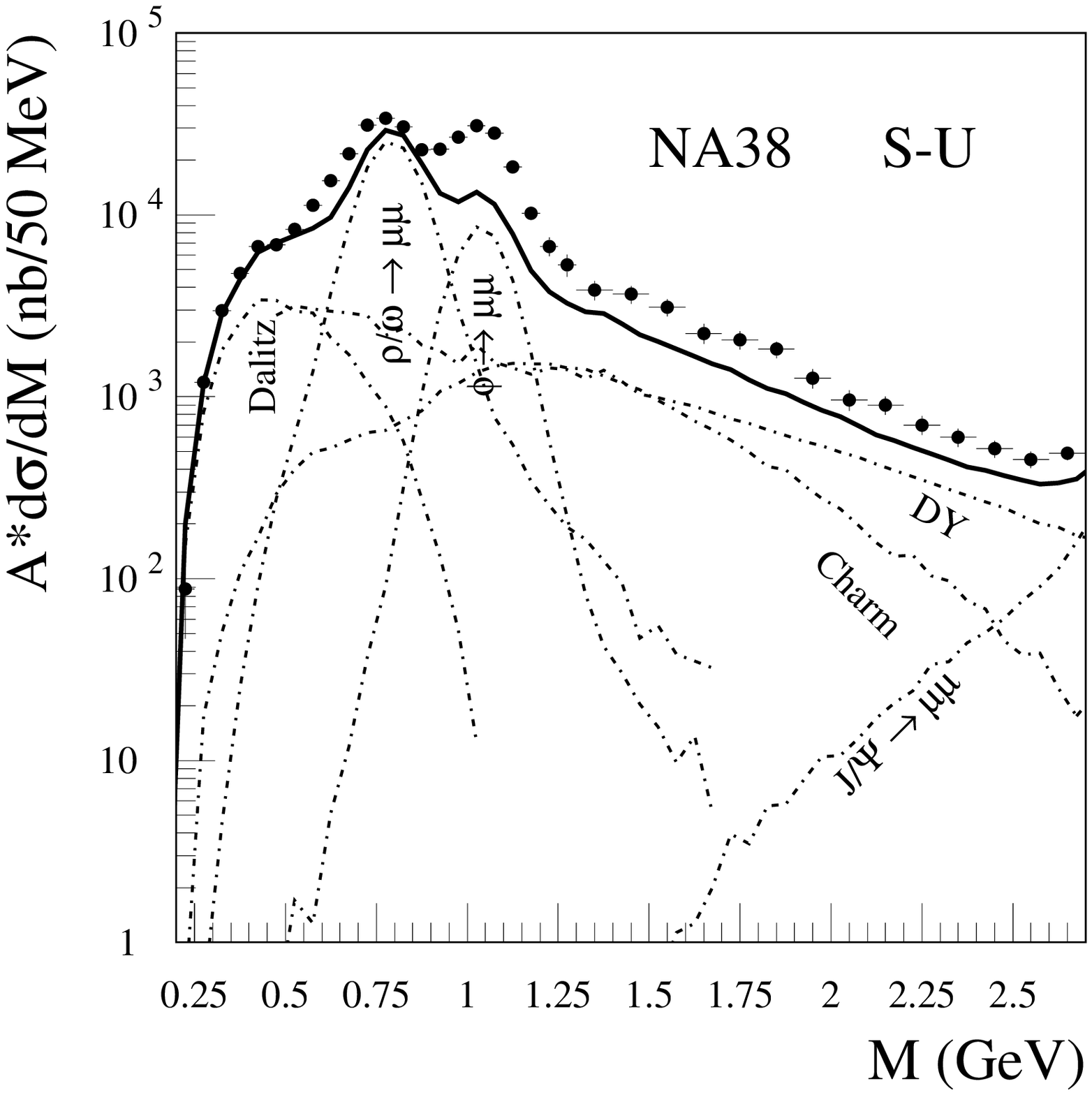,width=6.5cm,height=7.0cm}
\end{minipage}
\begin{minipage}[h!]{65mm}
\vspace*{-8.0cm}
\hspace*{6.5cm}
\epsfig{file=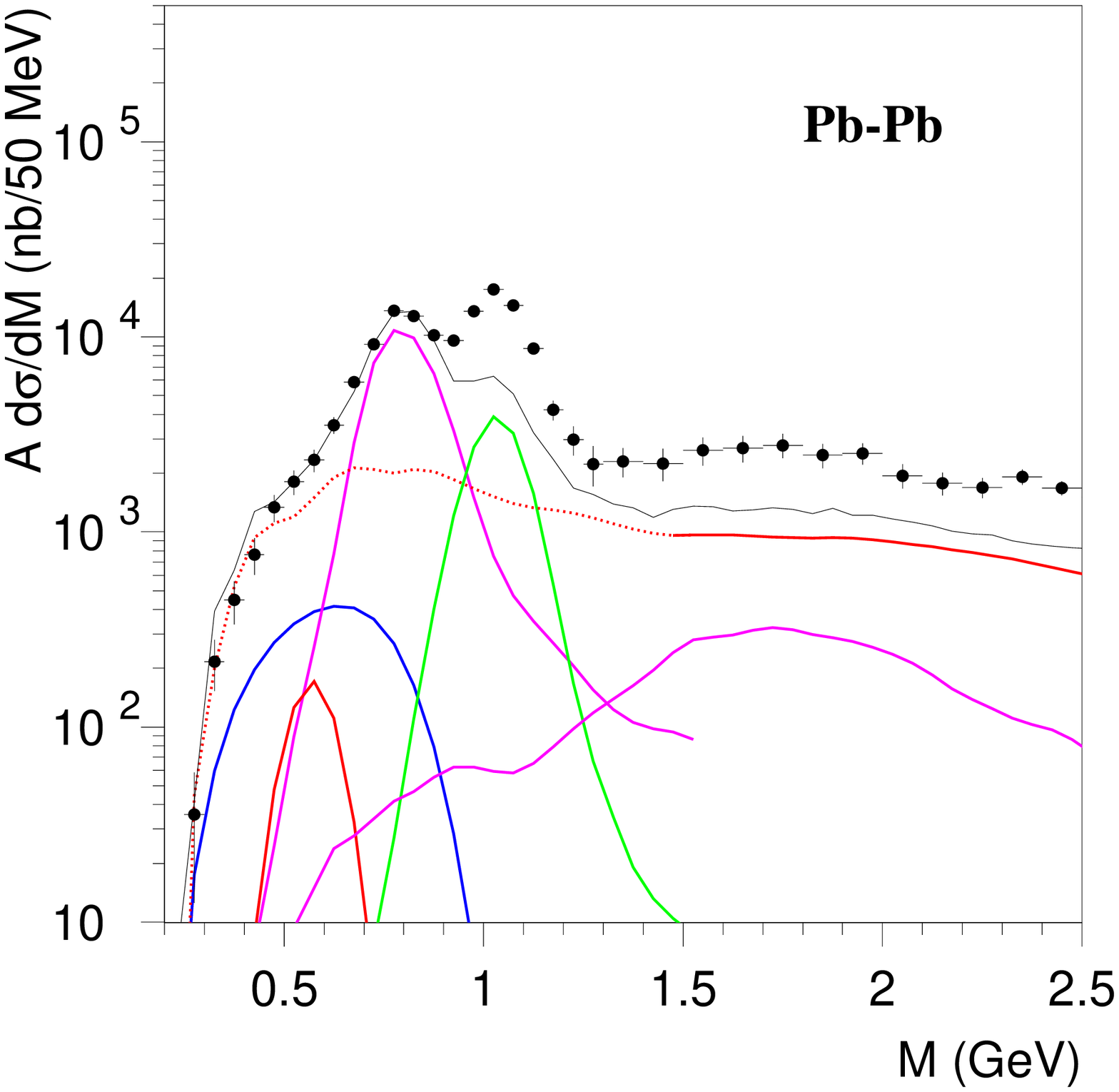,height=7.7cm,width=6.5cm}
\end{minipage}
%\vspace{-1.0cm}
\caption {Inclusive $\mu^+\mu^-$ mass spectra measured by NA38 in 
200 A~GeV S-U collisions (left panel) ~\protect\cite{falco-qm97} and by NA50 
in 158 A GeV Pb-Pb collisions (right panel) ~\protect\cite{bordalo-qm99}. The summed 
yield  of all known sources is shown as well as the individual contributions.}
%\vspace{-1.0cm}
\end{figure}
over the intermediate mass region.

There is a striking difference in the shape of the low-mass dilepton spectrum as 
measured by CERES and NA38/50. The latter exhibits a pronounced  structure 
due to the resonance decays whereas in the CERES results the 
structure is completely washed out (see Fig. 1).
Resolution effects are certainly not responsible for this difference
since the low-mass spectrum in p-Be and p-Au collisions 
measured by CERES with the same apparatus clearly shows the $\rho$/$\omega$ peak 
\cite{pbe-ee}. We also note that the two experiments cover nearly symmetric ranges 
around mid-rapidity ($\eta$~=~2.1~--~2.65 and $\eta$ = 3 -- 4 in CERES and NA38, 
respectively). 
But  CERES has a relatively 
low $p_t$ cut of 200 MeV/c on each track whereas NA38 is restricted to  
$m_t~>~$ 0.9~+~2($y_{lab} - 3.55)^2$ GeV/c$^2$. NA50 has an even stronger $m_t$ cut,
$m_t > $ 1.3 + 2($y_{lab} - 3.55)^2$ GeV/c$^2$. Moreover, NA38/50 have no centrality selection 
in the trigger whereas the CERES data corresponds to the top 30\% of the geometrical cross 
section. These two factors are likely to explain the apparent discrepancy since, as noted previously,
the excess observed by CERES is more pronounced at low pair p$_t$ and increases stronger than 
linearly with multiplicity.
Given enough statistics it should be fairly easy for the two experiments 
to apply common $m_t$ and centrality  cuts thereby allowing a 
direct and meaningful comparison between their results. Such a comparison should be possible once 
results from the newly approved experiment NA60 at CERN become available \cite{na60}.

\subsection{Theoretical Evaluation}
  The enhancement of low-mass dileptons has triggered a
wealth of theoretical activity. Dozens of articles have been published on the subject
and only a summary of the most prominent approaches and current open issues is presented here.
(For a comprehensive review see \cite{rapp-wambach}).  
 There is consensus that  a simple superposition of pp collisions
cannot explain the data and that an additional source is needed.
The pion annihilation channel ($\pi^+\pi^- \rightarrow \rho \rightarrow l^+l^-$), 
obviously not present in pp collisions, has to be added to the cocktail. This channel
provides first evidence of thermal radiation from a dense hadron gas and 
accounts for a large fraction of the observed enhancement. 
However. it is not sufficient to reproduce the data in the mass region  
0.2 $< m_{e^+e^-} <$ 0.6 GeV/c$^2$. These data have been quantitatively 
explained by taking into account in-medium modifications of the vector mesons.
Li, Ko and Brown \cite{li-ko-brown} were the first to propose a decrease of 
the $\rho$-meson mass in the hot and dense fireball as a precursor of chiral 
symmetry restoration, following the original Brown-Rho scaling \cite{brown-rho}. 
With this approach, an excellent agreement with the CERES data is achieved  
as demonstrated by the solid line (taken from \cite{rapp-qm99}) in Fig. 4.
%%% FIGURE 4 CERES 96 PB RESULTS WITH BR, RAPP AND KOCH. 
\begin{figure}[h!]
%\vspace{-2.0cm}
\centerline{\epsfxsize=8.5cm \epsfbox{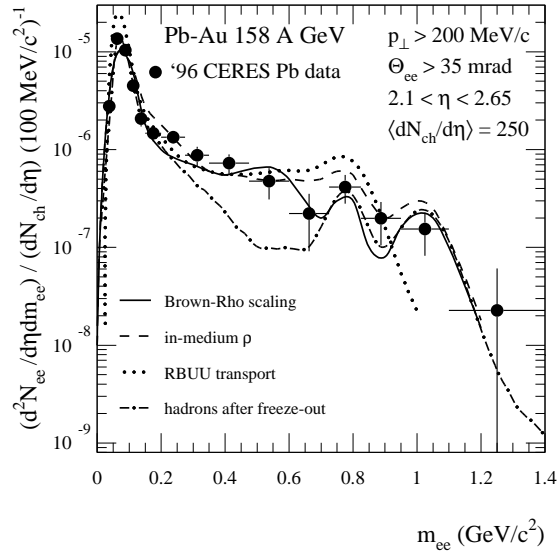}}
%\vspace{-1.8cm}
\caption {CERES results
         compared to calculations using dropping $\rho$ mass 
         (Brown-Rho scaling), in-medium $\rho$-meson broadening and RBUU transport model. 
         The dash-dotted line represents the yield from hadrons after freeze-out as in Fig. 1.   }
%\vspace{-0.5cm}
\end{figure}
 Another avenue, based on effective Lagrangians, uses a $\rho$-meson 
spectral function which takes into account the $\rho$ propagation in hot and dense
matter, including in particular the pion modification in the nuclear medium and 
the scattering of $\rho$ mesons off baryons \cite{wambach}. This leads to a large 
broadening of the $\rho$-meson line shape and consequently to a considerable 
enhancement of low-mass dileptons. These calculations achieve also an excellent reproduction
of the CERES  results as illustrated by the dashed line in Fig. 4.
Although the two approaches are different in the underlying physical picture 
(in the Brown-Rho scaling the constituent quarks are the relevant degrees of freedom 
whereas ref. \cite{wambach} relies on a hadronic description), it turns out that
the dilepton production rates calculated via hadronic and partonic models 
are very similar at SPS conditions, down to masses of about 0.5~GeV/c$^2$ 
\cite{rapp-wambach,rapp-qm99}. This ``quark-hadron duality''  explains the similar results
obtained with the two approaches. 

    Several issues remain still controversial. First, the Brown-Rho scaling hypothesis is not 
free of debate. Second,  both models, the dropping mass and the collision broadening of the  
$\rho$ meson, rely on a high baryon density. However, the role of baryons is still an open 
question. Calculations based on chiral reduction formulae, although similar in principle
to those of ref. \cite{wambach}, find very little effect due to baryons and are in fact 
low compared to the data \cite{zahed}.  The RBUU transport
calculations of Koch \cite{koch} find also very little effect due to the baryons and come to a 
reasonably close description of the data as shown in Fig. 4 by the dotted line. However,
the agreement is achieved by overestimating the  $\omega$ yield (and the  $\omega$ Dalitz decay yield)
as clearly seen in the figure in the mass region  m $\sim$ 800 MeV, which is dominated by the  
$\omega \rightarrow e^+e^-$ decay.
The CERES dilepton results in Pb-Au collisions at 40~GeV per nucleon (see Fig. 1 right panel)
could be very valuable in this debate. One of the incentives for this measurement was indeed
to study the effect of varying the baryon density which 
is expected to reach a maximum value, at least a factor of two larger than at 158 GeV per 
nucleon \cite{cassing}, at this low energy. 
Finally, there is a discrepancy between transport \cite{cassing-wambach}
and hydrodynamic calculations \cite{prakash} in treating the time evolution of the fireball,  
the former yielding a factor of 2-3 higher yields.
 
In a recent paper, Schneider and Weise argue that the CERES enhancement can be entirely explained by
thermal radiation emitted by the quark-gluon plasma through $\overline{q}q$ annihilation \cite{schneider-weise}. 
This is an interesting claim but also a surprising one since several previous studies came to the conclusion 
that the plasma phase shines too little with respect to the hadronic sources \cite{srivastava-sinha-gale96}.

\section{INTERMEDIATE MASS DILEPTONS}
    An  excess of dileptons has also  been observed in the intermediate mass region 
$1.5 < m < 3.0$ GeV/c$^2$ by  HELIOS-3 \cite{helios-3} and NA38/50  \cite{falco-qm97,bordalo-qm99} 
(see Fig. 3). The excess refers to the expected yield from
Drell-Yan and semi-leptonic charm decay which are the two main contributions 
in this mass region. The shape of the excess is very similar to the open charm contribution
and in fact doubling the latter nicely accounts for the excess. This is the basis for the
hypothesis of enhanced charm production put forward by NA38/50 \cite{bordalo-qm99}. 
However it seems unlikely that at 
SPS energies charm production could be enhanced by such a large factor \cite{shor}. HELIOS-3
points into a different direction. The excess plotted as a function of the dimuon transverse
mass can be fitted by a single exponential shape below and above the resonance decays \cite{helios-3},  
suggesting a common origin of the excess in the low and intermediate mass
regions. Following this line, Li and Gale \cite{li-gale}
calculated the invariant dimuon spectrum in 
central S-W collisions at 200 A GeV. On top of the {\it physics}
%%%%%% FIGURE 5.  FROM RAPP AND SHURYAK ON NA50
\begin{figure}[t]
%\vspace{-1.0cm}
\begin{minipage}[t]{65mm}
\epsfig{file=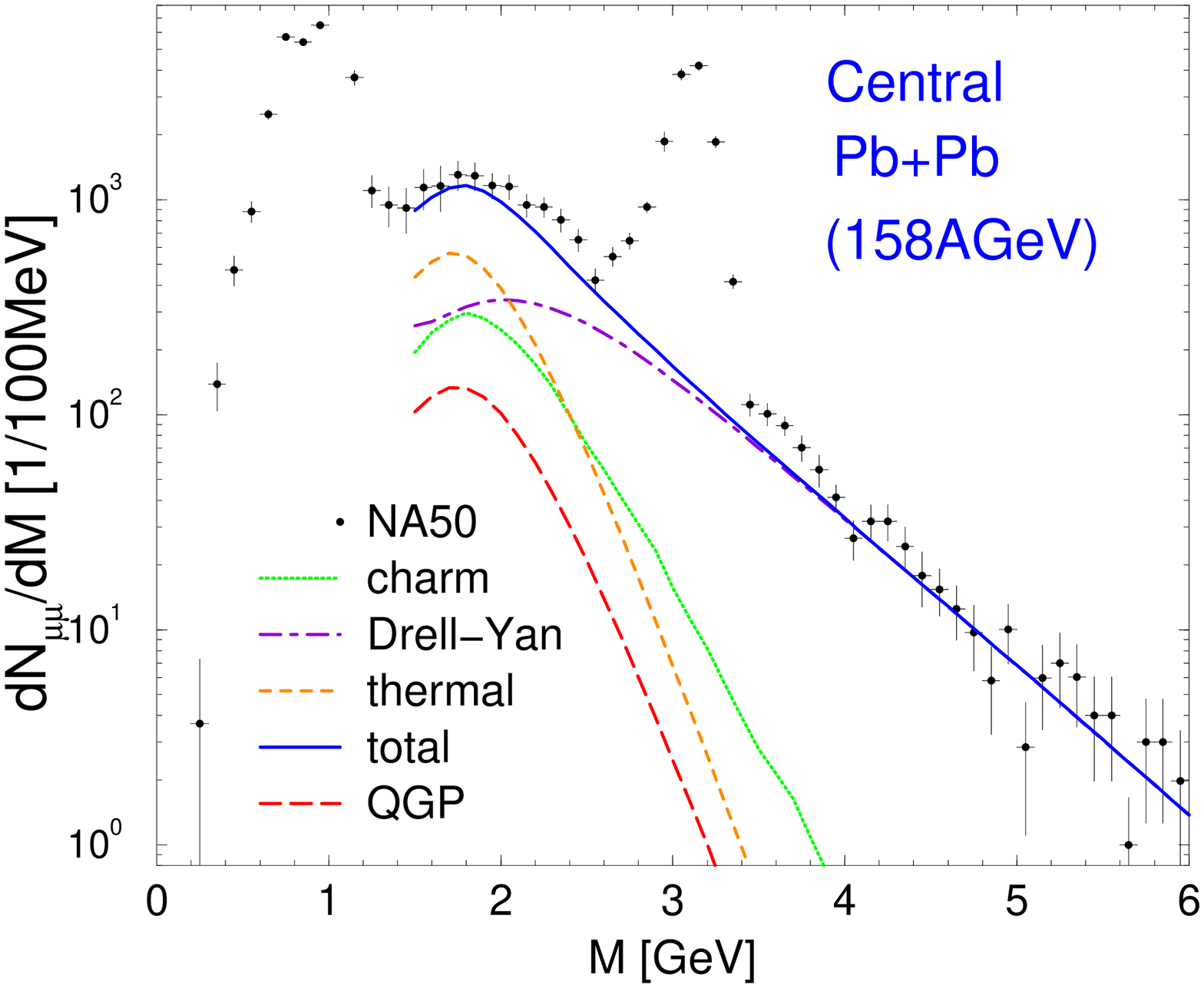,width=6.5cm,height=6.5cm,angle=-90}
\end{minipage}
\begin{minipage}[h!]{65mm} 
\vspace*{-6.5cm}
\hspace*{6.5cm}
\epsfig{file=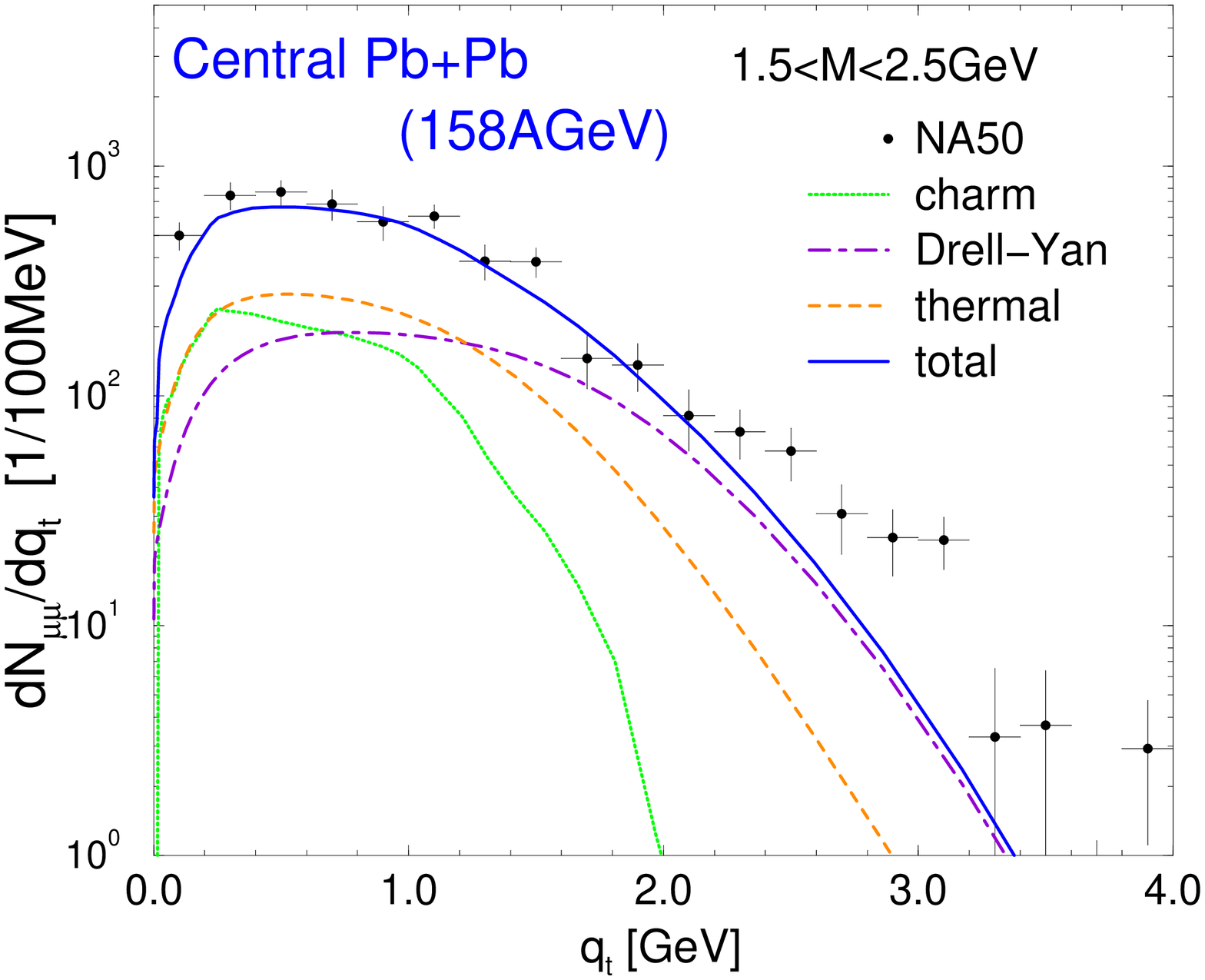,height=6.5cm,width=6.5cm,angle=-90}
\end{minipage}
%\vspace{-1.5cm}
\caption{NA50 dimuon data (left panel: invariant mass spectrum, right panel: transverse momentum spectrum)
         compared  to calculations of Rapp and Shuryak~\protect\cite{rapp-shuryak}. The various contributions
         (open charm, Drell Yan, thermal radiation) and their sum are indicated in the figures.}
\end{figure}
background of Drell-Yan and open charm pairs, they considered the thermal radiation of muon pairs
resulting from secondary meson interactions including higher resonances and in particular 
the $\pi a_1 \rightarrow l^+l^-$. The calculations are based on the same relativistic 
fireball model that successfully reproduces the low-mass dileptons discussed in the previous 
section \cite{li-ko-brown}. Whereas they could reproduce the low-mass data only with the dropping mass
scenario, in the intermediate mass region the difference between free and in-medium
meson masses with respect to the data is not so large. The calculations with
free masses slightly overestimate the data whereas with dropping masses 
the situation is reversed.  The intermediate mass region alone cannot therefore be used to 
support the dropping mass model, however it is important that the model can 
simultaneously explain the low and intermediate mass regions.

   Recent calculations \cite{rapp-shuryak,kampfer} were also able to explain the NA50 
enhancement in the intermediate mass region as originating from thermal radiation.
Rapp and Shuryak \cite{rapp-shuryak} exploit the quark-hadron duality and use the $q\overline{q}$
annihilation rates to calculate the dilepton yield throughout the entire space-time 
evolution of the collision which they describe with a simple fireball model. The thermal radiation, 
added to the ``background'' (Drell Yand and open charm) contributions, reproduces the NA50 
enhancement observed in the mass range $1.5 < m < 3.0$  GeV/c$^2$ as shown in Fig. 5 (left panel).
Only a small fraction of this radiation is emitted at the early stages and is associated 
with the QGP phase. Their calculations also reproduce the transverse momentum dependence of the
muon pairs in the same mass interval (see Fig. 5 right panel).
 
\section{DIRECT PHOTONS}

   Direct photons are expected  to provide analogous information 
to thermal dileptons since real and virtual photons should carry the same physics
information. However, the physics background for real photons (mainly 
from $\pi^0$ and $\eta$ decays) is larger
by orders of magnitude compared to dileptons (at m $>$ 200 MeV/c$^2$), making the measurement of photons much less
sensitive to a new source. And indeed,  in contrast with the dilepton results, there is 
no clear evidence of enhancement
in the measurements of real photons. All experiments performed with O and S 
beams have been able to establish only an upper limit for the production of 
thermal photons, which is of the order of 10 -- 15\%  of the expected yield 
from hadron decays. The sensitivity is actually limited not by statistics but by
the systematic errors. 
The S-Au results have been quantitatively reproduced \cite{li-brown}  by the same 
fireball model, including dropping masses, used to explain the 
CERES and HELIOS-3 dilepton results.  The calculations predict an excess
of direct photons with respect to the hadronic background of a few
percent, in agreement with the experimental results 
and with a simple estimate based on order of magnitude considerations \cite{it-qm95}.
%%%%%% FIGURE 6. WA98 PHOTONS DATA
\begin{figure}[h!]
%\hspace*{6.5cm}
\centerline{\epsfxsize=8.5cm \epsfbox{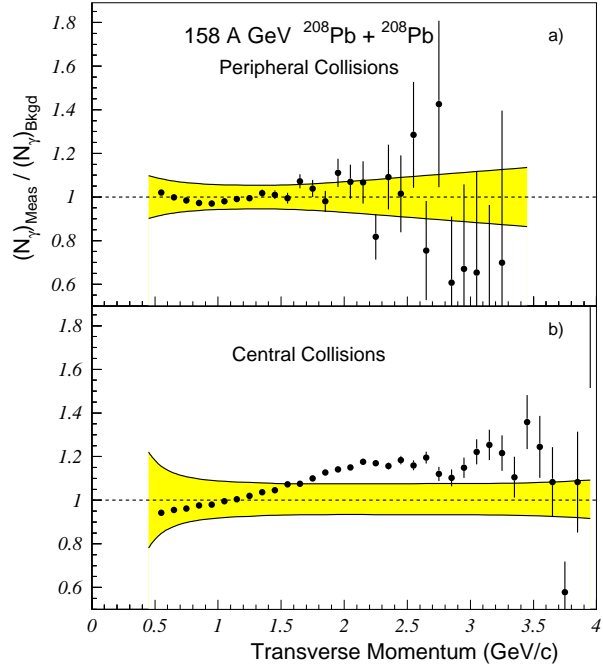}}
%\vspace{-1.5cm} 
\caption{WA98 photon data in Pb-Pb peripheral (top panel) and central (lower panel) collisions at 
158 A GeV. The error bars represent the statistical errors whereas the shaded bands indicate the 
systematic errors ~\protect\cite{wa98-pb-ph}.}
%\vspace{-0.8cm}
\end{figure}
In Pb-Pb collisions at 158 A GeV a somewhat larger effect has been reported by WA98~\cite{wa98-pb-ph}. 
In central collisions, a direct photon yield of $\sim20\%$ is observed at p$_t~ >$~1.5GeV/c
whereas for peripheral collisions no excess is observed, as illustrated in Fig. 6 where 
the ratio of the measured photons to the calculated photons from hadronic sources is plotted 
as a function of p$_t$. Given the errors as indicated in the figure, the direct photon 
yield is only 1$\sigma$ effect.  

    The p$_t$ spectrum of the direct photons, defined as the difference between the measured photon
and the hadronic background, is shown in Fig. 7. The data of S-Au collisions scaled by a factor 
of 3.5 \cite{dinesh-photons} are also included. The error bars represent statistical and systematic 
errors added in quadrature. Data points with downward 
%%%%%% FIGURE 7. WA98 PHOTONS DATA WITH SRIVASTAVA'S CALCULATIONS
\begin{figure}[h!]
%\hspace*{6.5cm}
\centerline{\epsfxsize=8.5cm \epsfbox{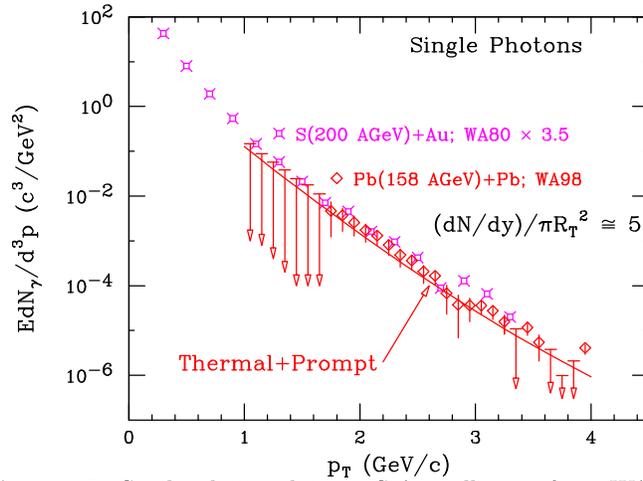}}
%\vspace{-1.5cm} 
\caption{Single photon data in S-Au collisions from WA80 experiment scaled by a factor of 3.5
and in Pb-Pb central collisions  from WA98 experiment compared to  calculations from ref.
~\protect\cite{dinesh-photons}.}
%\vspace{-0.8cm}
\end{figure}
errors represent the 90\%CL upper limit of 
the direct yield. The figure also displays results of calculations assuming the initial 
formation of a quark-gluon plasma, which expands, cools down, hadronizes and finally 
freezes out \cite{dinesh-photons}. The calculations  describe
the data very well with the sum of almost equal contributions from prompt and thermal 
photons. The latter includes photons from the deconfined phase (calculated with the 
corrected two-loop rates \cite{aurenche}) and the hadron phase. At low p$_t$ the quark 
matter contribution is relatively small whereas at high p$_t$ it appears to be the dominant source
of thermal radiation.

\section{RHIC OPPORTUNITIES}
   RHIC will allow the study of Au-Au collisions up to a maximum  energy
of $\sqrt{s_{NN}}=200$ GeV. This unprecedented energy, more than one order 
of magnitude higher than at the SPS, will offer the possibility to extend  the studies of 
quark matter under much better conditions.  
   The higher collision energies will result in higher initial temperatures  
or equivalently in higher initial energy densities.  This will lead to longer lifetimes 
and larger volumes allowing for a better study of the system before it freezes out 
and in particular of the early stages. 
    In addition, the ability of RHIC to produce beams of different energies and species 
guaranties a systematic and comprehensive study of nuclear collisions from the elementary 
pp collisions to  p-nucleus, and from light nuclei up to gold-gold collisions.

   The first RHIC run on Au-Au collisions at   $\sqrt{s_{NN}}=130$ GeV  took place 
in the summer of year 2000. From a short and relatively very low luminosity run,  
the four experiments were able to obtain an impressive 
amount of results, mainly in the hadronic sector, covering global observables, negative, 
positive and identified particle spectra, particle ratios, two particle correlations, flow 
studies and more. The results were presented  at the Quark Matter Conference which took place at Stony Brook 
a few months after the RHIC year-1 data taking \cite{qm01} and they allow 
to make a quantitative comparison
of the main characteristics of central nuclear collisions at RHIC relative to the SPS.
Some of them are listed in Table 2. Both the charged particle rapidity density and the
transverse energy are $\sim 70-80\%$ larger at RHIC than at SPS. Using the Bjorken 
prescription \cite{bjorken}, this translates into a similar increase of the energy density.

\begin{table}[htb]
%\vspace{-1.0cm}
\medskip
\begin{center}
{Table 2. Conditions at RHIC year-1 running relative to SPS }\\[0.3cm]
\renewcommand{\arraystretch}{1.5}
%\begin{tabular}{||c|c|c||} \hline\hline
\hskip2pc\vbox{\columnwidth=8cm
\begin{tabular}{|c|c|c|}
% \hline\hline
                                      & RHIC          &  SPS             \\  
                                      & Au + Au       &  Pb + Pb         \\ \hline
$\sqrt{s_{NN}}$ (GeV)                 & 130           & 17.2              \\ \hline
dN$_{ch}$/dy (y=0)  $^{a)}$           & 750           & 410               \\ \hline
dE$_{t}$/dy (y=0) [GeV]               & 688  $^{b)}$  & 405  $^{c)}$      \\ \hline
$\epsilon [GeV/fm^3]$  $^{b)}$        & 5.0           &  2.9         \\ \hline
$\overline{p}/p$  (y=0)               & 0.60 $^{d)}$  & 0.07  $^{e)}$     \\ \hline
%\hline \hline 
\end{tabular}
}
\end{center}
a) see ref.  \cite{phenix-ppg1} for details of the quoted values.\\
b) see ref.  \cite{phenix-ppg2} for details of the quoted values.\\
c) see ref. \cite{na49-et}.\\ 
d) see refs. \cite{phobos-ppbar}.\\ 
e) see ref. \cite{na49-qm99}.
\end{table}
 
   The low luminosity of RHIC year-1 run is clearly insufficient to address any of
the physics topics discussed in this paper. However, from the unique features 
of RHIC one can expect drastic differences. As discussed above, the CERN results 
provide evidence of thermal radiation from a dense hadron gas whereas no
clearly convincing evidence was found of electromagnetic radiation directly emitted from the 
QGP. The improved conditions at RHIC should facilitate the 
direct observation of this radiation thus providing a key element in the quest for 
quark-gluon plasma formation in nuclear collisions. The radiation should be observable in 
principle in the direct photon channel and in the dilepton channel. At initial temperatures 
likely to be reached at RHIC, theory has singled out the dilepton mass range of 1-3 GeV/c$^2$ 
as the most suitable window to observe the thermal radiation from the QGP 
phase~\cite{kajantie86,ruuskanen92}.

  Another interesting difference might be in the role of baryons. Theory explains the low-mass 
pair enhancement observed at CERN by invoking in-medium modifications 
of the $\rho$ meson properties as precursor of chiral symmetry restoration and,
as discussed in Section 2.2,  baryon density is 
the major factor responsible for these modifications both in the dropping mass and in the collision
broadening scenarios. 
  At RHIC, conditions of higher charged particle densities and hence higher temperatures
may offer the possibility to explore a new domain where temperature rather than baryon density 
would be the dominant factor.  The first RHIC results show that the net-baryon density is indeed 
much smaller than at the SPS, although it is clearly not yet net-baryon free. As shown 
in Table 2, the $\overline{p}/p$ ratio increases from $\sim$0.07 at the SPS \cite{na49-qm99} 
to 0.60-0.65 at RHIC
\cite{phobos-ppbar,star-ppbar}. However, the relevant factor is not the net baryon but the total 
baryon density and if there is substantial baryon/antibaryon production at RHIC 
(no absolute values have been reported yet) the role of baryons could still be a significant one.
 
    Among the four different experiments at RHIC, the PHENIX experiment is particularly
focussed on the measurement of electromagnetic probes and is expected to address all these issues.
First results should become available with the expected higher luminosity of RHIC year-2
running.


\begin{thebibliography}{99}
\bibitem{qm} See e.g. the Proceedings of Quark Matter 99 published in  
      Nucl. Phys. {\bf A661}.

\bibitem{pbe-ee} G. Agakichiev et al., CERES Collaboration,
      Euro. Phys. Jour. {\bf C4}, 231 (1998). 
     
\bibitem{pbe-eegamma} G. Agakichiev et al., CERES  Collaboration,
      Euro. Phys. Jour. {\bf C4}, 249 (1998). 

\bibitem{prl95} G. Agakichiev et al.,  CERES Collaboration,
       Phys. Rev. Lett. {\bf 75}, 1272 (1995).

\bibitem{ir-qm97} I. Ravinovich, for the CERES Collaboration, 
      Nucl. Phys. {\bf A638}, 159c (1998).

\bibitem{pbau95-plb} G. Agakichiev et al., CERES Collaboration,
       Phys. Lett. {\bf B422}, 405 (1998).

\bibitem{bl-qm99} B. Lenkeit, for the CERES Collaboration, 
      Nucl. Phys. {\bf A661}, 23 (1999).

\bibitem{pb-au40} G. Agakichiev et al.,  CERES Collaboration,
       Nucl. Phys. {\bf A}, in press.

\bibitem{helios-1} T. Akesson et al., HELIOS-1 Collaboration,	
       Z. Phys. {\bf C68}, 47 (1995).

\bibitem{helios-3} M. Masera for the HELIOS-3 Collaboration,
       Nucl. Phys. {\bf A590}, 93c (1995) and A.L.S. Angelis et. al., HELIOS-3 Collaboration,	
       CERN-EP/98-82.

\bibitem{falco-qm97} A. De Falco, for the NA38 Collaboration, 
       Nucl. Phys. {\bf A638}, 487c (1998).

\bibitem{bordalo-qm99} E. Scomparin et al., NA50 Collaboration,
       J. Phys.  {\bf G25}, 235 (1999) and
       P. Bordalo, for the NA50 Collaboration,
       Nucl. Phys. {\bf A661}, 638 (1999).

\bibitem{ceres-ph} R. Baur et al., CERES Collaboration,
       Z. Phys. {\bf C71}, 571 (1996).

\bibitem{helios2-ph} T. Akesson et al., HELIOS-2 Collaboration, 
       Z. Phys. {\bf C46}, 369 (1990).
    
\bibitem{wa80-O-ph} R. Albrecht et. al., WA80 collaboration, 
       Z. Phys. {\bf C51}, 1 (1991).

\bibitem{wa80-S-ph} R. Albrecht et al., WA80 Collaboration,
       Phys. Rev. Lett.  {\bf 76}, 3506 (1996).
       
\bibitem{wa98-pb-ph} M.M. Aggarwal et al., WA/98 Collaboration,
       Phys. Rev. Lett.  {\bf 85}, 3595 (2000).


\bibitem{pbm-js} P. Braun-Munzinger, I. Heppe and J. Stachel, 
        Phys. Lett. {\bf B465}, 15 (1999).

\bibitem{na60} NA60 Proposal to CERN/SPSC 2000-010, SPSC / P 316, March 2000.

\bibitem{rapp-wambach} R. Rapp and J. Wambach,
       Preprint hep-ph/9909229 to appear in Adv. Nucl. Phys.

\bibitem{li-ko-brown} G.Q. Li, C.M. Ko and G.E. Brown,
        Phys. Rev. Lett. {\bf 75}, 4007 (1995).
 
\bibitem{brown-rho} G.E. Brown and M. Rho,
        Phys. Rev. Lett. {\bf 66}, 2720 (1991) and 
        Phys. Rep.{\bf 269}, 333 (1996).  

\bibitem{rapp-qm99} R. Rapp, 
       Nucl. Phys. {\bf A661}, 33 (1999).

\bibitem{wambach} R. Rapp, G. Chanfray and J. Wambach,
       Nucl. Phys. {\bf A617}, 472 (1997).
       and J. Wambach, 
       Nucl. Phys. {\bf A638}, 171c (1998).

\bibitem{zahed} J.V. Steele, H. Yamagishi and I. Zahed,
      Phys. Rev {\bf D56}, 5605 (1997). 
      and J.V. Steele and I. Zahed, 
      Phys. Rev {\bf D60}, 037502 (1999).  

\bibitem{koch} V. Koch, nucl-th/9903008 and
      V. Koch and C. Song,
      Phys. Rev. {\bf C54}, 1903 (1996).

\bibitem{cassing} E.L. Bratkovskaya and W. Cassing,
      Nucl. Phys. {\bf A619}, 413 (1997).

\bibitem{cassing-wambach}W. Cassing, E.L. Bratkovskaya, R. Rapp, J. Wambach,
       Phys. Rev {\bf C57}, 916 (1998).

\bibitem{prakash} P. Huovinen and M. Prakash,
      Phys. Lett.  {\bf B450}, 15 (1999). 

\bibitem{schneider-weise} R.A. Schneider and W. Weise,
      Eur. Phys. J. {\bf A9}, 357 (2000).

\bibitem{srivastava-sinha-gale96} See e.g. D.K. Srivastava, B. Sinha and C. Gale,
      Phys. Rev {\bf C53}, R567 (1996), 
      and R. Rapp and J. Wambach,
      Eur. Phys. J. {\bf A6}, 415 (1999).

\bibitem{shor} A. Shor,
      Phys. Lett.  {\bf B215}, 375 (1988). 

\bibitem{li-gale} G.Q. Li and C. Gale,
      Phys. Rev. Lett. {\bf 81}, 1572 (1998). 

\bibitem{rapp-shuryak} R. Rapp and E. Shuryak, 
       Phys. Lett.  {\bf B473}, 13 (2000). 

\bibitem{kampfer} K. Gallmeister, B. Kampfer and O.P. Pavlenko,
       Phys. Lett.  {\bf B473}, 20 (2000). 

\bibitem{li-brown} G.Q. Li and G.E. Brown,
       Nucl. Phys. {\bf A632}, 153 (1998). 

\bibitem{it-qm95} I. Tserruya,
        Nucl. Phys. {\bf A590}, 127c (1995).

\bibitem{dinesh-photons} D.K. Srivastava,
      Preprint nucl-th/0102005 and
      D.K. Srivastava and B. Sinha,
      Preprint nucl-th/0006018.

\bibitem{aurenche} F.D. Steffen and M.H. Thoma, 
      Preprint hep-ph/0103044.

\bibitem{qm01} Proc. of Quark Matter 2001, to be published in 
      Nucl. Phys. {\bf A}.

\bibitem{phenix-ppg1} K. Adcox et al., the PHENIX Collaboration,
      Phys. Rev. Lett. {\bf 86}, 3500 (2001). 

\bibitem{phenix-ppg2} K. Adcox et al., the PHENIX Collaboration,
      Preprint nucl-ex/0104015. 

\bibitem{na49-et} T. Alber et al., NA49 Collaboration,
      Phys. Rev. Lett. {\bf 75}, 3814 (1995). 

\bibitem{phobos-ppbar} B.B. Back et al., the PHOBOS Collaboration, 
       hep-ex/0104032.

\bibitem{na49-qm99} F. Sickler et al., NA49 Collaboration,
      Nucl. Phys. {\bf A661}, 45cc (1999).

\bibitem{bjorken} J.D. Bjorken,
      Phys. Rev. {\bf D27}, 140 (1983).
 
\bibitem{kajantie86} K. Kajantie, J. Kapusta, L. McLerran and A. Mekjian,
       Phys. Rev. {\bf D34}, 2746 (1986).

\bibitem{ruuskanen92} P.V. Ruuskanen, 
       Nucl. Phys. {\bf A544}, 169c (1992).

  
\bibitem{star-ppbar}  C. Adler et al., the STAR Collaboration,
      nucl-ex/0104022.


\end{thebibliography}
\end{document}